\documentclass{aa}

\usepackage[dvips]{graphicx}
\def\ppt#1{\frac{\partial#1}{\partial t}}
\def\diver#1{\nabla\!\cdot\!#1}

\newcommand{\bu}{{\rm B}_{{\rm u}}}
\newcommand{\buz}{{\rm B}_{{\rm u},0}}
\newcommand{\br}{\delta{\rm B}}

\newcommand{\btot}{{\rm B}_{{\rm tot}}}
\newcommand{\temp}{{\rm T}}
\newcommand{\pre}{{\rm P}}

\newcommand{\vrms}{{\rm V}_{{\rm rms}}}
\begin{document}

%
 \title{Global Dynamical Evolution of the ISM in Star Forming Galaxies}
 \subtitle{I. High Resolution 3D HD and MHD Simulations: Effect of the Magnetic Field}

   \author{Miguel A. de Avillez
          \inst{1}
          \and
          Dieter Breitschwerdt\inst{2}
          }
   \offprints{M. A. de Avillez}

   \institute{Department of Mathematics, University of \'Evora,
              R. Rom\~ao Ramalho 59, 7000 \'Evora, Portugal \\
              email: mavillez@galaxy.lca.uevora.pt
     \and
        Institut f\"ur Astronomie, Universit\"at Wien,
        T\"urkenschanzstr. 17, A-1180 Wien, Austria\\
        email: breitschwerdt@astro.univie.ac.at
            }

   \date{Received ** October 2004/ Accepted 31 January 2005}

   \titlerunning{High resolution HD and MHD Simulations of the ISM}

   \abstract{In star forming disk galaxies, matter circulation between
   stars and the interstellar gas, and, in particular the energy input by
   random and clustered supernova explosions, determine the dynamical
   and chemical evolution of the ISM, and hence of the galaxy as a
   whole. Using a 3D MHD code with adaptive mesh refinement developed
   for this purpose, we have investigated the r\^ole of magnetized
   matter circulation between the gaseous disk and the surrounding
   galactic halo. Special emphasis has been put on the effect of the
   magnetic field with respect to the volume and mass fractions of the
   different ISM ``phases'', the relative importance of ram, thermal
   and magnetic pressures, and whether the field can prevent matter
   transport from the disk into the halo. The simulations were
   performed on a grid with an area of 1 kpc$^{2}$, centered on the
   solar circle, extending $\pm 10$ kpc perpendicular to the galactic
   disk with a resolution as high as 1.25 pc. The simulations were run
   for a time scale of 400 Myr, sufficiently long to avoid memory
   effects of the initial setup, and to allow for a global dynamical
   equilibrium to be reached in case of a constant energy input rate.
   The main results of our simulations are: (i) The $\temp \leq
   10^{3}$ K gas is mainly concentrated in shock compressed layers,
   exhibiting the presence of high density clouds with sizes of a few
   parsecs and $\temp \leq 200$ K. These structures are formed in
   regions where several large scale streams of convergent flow
   (driven by SNe) occur. They have lifetimes of a few free-fall
   times, are filamentary in structure, tend to be aligned with the
   local field and are associated with the highest field strengths;
   (ii) the magnetic field has a high variability and it is
   \emph{largely uncorrelated with the density}, suggesting that it is
   driven by superalfvenic inertial motions; (iii) ram pressure controls the flow
   for $200<\temp\leq 10^{5.5}$ K. For $\temp \leq 200$ K magnetic
   pressure dominates, while the hot gas ($\temp > 10^{5.5}$ K) in
   contrast is controlled by the thermal pressure, since magnetic
   field lines are swept towards the dense compressed walls; (iv) up
   to $49\%$ of the mass in the disk is concentrated in the classical
   thermally \emph{unstable} regime $200<\temp\leq 10^{3.9}$ K with
   $\sim 65\%$ of the warm neutral medium (WNM) mass enclosed in the
   $500\leq \temp \leq 5000$ K gas, consistent with recent
   observations; (v) the volume filling
   factors of the different temperature regimes depend sensitively on
   the existence of the duty cycle between the disk and halo, acting
   as a pressure release mechanism for the hot phase in the disk. We
   find that in general gas transport into the halo in 3D is not
   prevented by an initial disk parallel magnetic field, but only
   delayed initially, for as long as it is needed to punch holes into
   the thick magnetized gas disk.  The mean volume filling factor of
   the hot phase in the disk is similar in HD and MHD (the latter with
   a total field strength of 4.4 $\mu$G) runs, amounting to $\sim
   17-21\%$ for the Galactic supernova rate.
  }
\maketitle


\section{Introduction}
\label{intro}
The evolution of galaxies is largely dependent on the physical and
chemical state of the interstellar medium and thus on the rate at
which they can form stars. This immediately raises the question of the
structure of the ISM, i.e.\ the distribution of gas in so-called
phases, thought to be stable regions in the $p-V$-diagram, with
respect to entropy perturbations. A detailed inventory and estimate of
timescales of heating and cooling processes suggested the presence of
a two (Field et al. 1969) -- and after the widespread interstellar
O{\sc vi} had been discovered -- a three phase ISM (e.g.\ McKee \&
Ostriker 1977), consisting of a cold, warm and hot component in
pressure equilibrium. Since the hot intercloud medium is pervasive, it
was deemed possible that it could fill a galactic corona as had been
suggested by Spitzer (1956) in order to confine high velocity
clouds. Observationally it was difficult to determine reliably volume
filling factors for these phases in our Galaxy due to the
observational vantage point. However, it became evident that the small
surface coverage of H{\sc i} holes in external galaxies (e.g.\ Brinks
\& Bajaja 1986) argues for a much lower volume filling factor of the
hot phase there. Since at the same time the extended H{\sc i} (Lockman
1984) and H{\sc ii} (Reynolds 1985) layers of the Milky Way were
discovered, it seemed plausible that break-out of supernova remnants
(SNRs) was inhibited (unless they occurred at a significant height
above the disk) and only the most energetic superbubbles (SBs) with at
least 800 SNe in concert (see Koo \& McKee 1992) would achieve
blow-out of the disk.  Things might even become worse, once a disk
parallel magnetic field is considered, which according to observations
should have a regular component of the order of $3 \, \mu$G or even
higher (s.~Beck 2004).  If the scale height of the field is infinite,
then bubbles should stall according to Tomisaka (1998) already for a
moderate stellar OB association of about 50 members.

On the other hand, owing to the high sensitivity and large
throughput of the ROSAT XRT (Tr\"umper 1983) and its PSPC
instrument, a number of normal spiral galaxies with soft X-ray halos
were detected, e.g.\ \object{NGC 891} (Bregman \& Pildis 1994) and
\object{NGC 4631} (Wang et al. 1995, Vogler \& Pietsch 1996). In
some cases, even local correlations between H$\alpha$, radio
continuum and soft X-rays were found (see Dettmar 1992), arguing for
local outflows as it had been suggested in the Galactic fountain
(Kahn 1981), chimney (Norman \& Ikeuchi 1989) and the Galactic wind
model (Breitschwerdt et al. 1991, Breitschwerdt \& Schmutzler 1994).
However, until a few years ago it was not possible to undertake an
extended 3D numerical study in order to follow the evolution of ISM
structures on small and large scales simultaneously, as was done
for the fountain by Avillez (1998). Recently, Korpi et al. (1999)
have numerically studied the evolution of superbubbles in a
magnetized disk and found that blow-out is much more likely than
previously thought, mainly because bubbles evolve in an
inhomogeneous background medium. Their 3D simulations are however
limited by the usage of a small grid (1~kpc perpendicular to the
disk), so that they were able to follow only the onset of a
supernova disturbed ISM for a short timescale of $\leq 100$ Myr. As
a consequence,  these results still bear the imprint of the initial
conditions, and the development of a disk-halo cycle could not be
followed.

It is a fortunate coincidence that exactly 100 years after the discovery
of the ISM by stationary Ca{\sc ii} lines (Hartmann 1904), we are
now for the first time in the position to give a detailed picture of
the structure and evolution of the ISM on the scale of parsecs and below, that seems to converge (i.e.,
does not change substantially by merely increasing the numerical
resolution). Our contribution here is to have carried out large scale
high resolution 3D simulations of the ISM that include the Galactic
magnetic field, background heating due to starlight and allow for
the establishment of the duty-cycle between the disk and halo
(commonly known as galactic fountain) by using a grid that extends
up to 10 kpc on either side of the midplane. Our simulations capture
both the largest structures (e.g., superbubbles) together with the
smaller ones (e.g., filaments and eddies) down to 1.25 pc.
We investigate, among other things, the variability
of the magnetic field in the Galactic disk and its correlation with
the density, the r\^ole of ram pressure in the dynamics of disk gas
and the relative weight of the ram, thermal and magnetic pressures,
the mass distribution and the volume filling factors of the
different temperature regimes in the ISM.  Other important issues
like the variation of the volume filling factors of the ISM
''phases'' with energy injection rate by SNe, the dynamics of the
galactic fountain, the conditions for dynamical equilibrium and the
importance of convergence of these results with increasing grid
resolution have been treated in Avillez (2000) and Avillez \&
Breitschwerdt (2004, hereafter Paper~I).

The outline of the present paper is as follows: in Section~2 the
model and numerical setup used in the simulations is presented;
Section~3 deals with the most important results obtained in these
MHD simulations, their interpretation, and comparison with those in
corresponding HD runs described in Paper~I. In Section~4 a
discussion of the results and comparison with other models is
carried out. Section~5 closes the paper with a summary of the main
results and some final remarks on future work.

\section{Model and Simulations}

We have run high resolution kpc-scale MHD simulations
of the ISM, driven by SNe (with a canonical explosion energy of
$10^{51}$ erg) at the Galactic rate, on a Cartesian grid of $0\leq
(x,y)\leq 1$ kpc size in the Galactic plane and extending $-10\leq z
\leq 10$ kpc into the halo. These runs use a modified version of the model
described in Avillez (2000) coupled to a three-dimensional numerical
AMR scheme that uses the MHD extension of the Piecewise Parabolic
Method (Dai \& Woodward 1994, 1998) and the grid refinement procedure
discussed in Balsara (2001) assuring that {\bf B} is divergence-free during the refinement process and simulation
time. The complete set of MHD equations that are solved numerically
can be written in the conservative form
\begin{eqnarray}
\ppt{\rho} + \diver{{\bf \Pi}} =  S_{\rho} & & \mbox{(conservation of mass)},
\label{mass}\\
\ppt{{\bf \Pi}} + \diver{{\bf T}} =  S_{\Pi} && \mbox{(conservation of momentum)},
\label{momentum}\\
\ppt{E} + \diver{{\bf U}} =  S_{E} & & \mbox{(conservation of energy)},
\label{energy}\\
\ppt{{\bf B}} + \diver{{\bf Y}} =  0 & &\mbox{(conservation of magnetic flux)},
\label{bfield}
\end{eqnarray}
where ${\bf \Pi}=\rho {\bf u}$ is the mass flux, ${\bf
T}=\rho {\bf u}{\bf u}+\left(P_{th}+\frac{1}{2}B^{2}\right){\bf
I}-{\bf B}{\bf B}$ is the stress tensor, $E=\frac{1}{2}\rho
u^{2}+\rho\Phi+\frac{1}{\gamma-1}P_{th}+\frac{1}{2}B^{2}$ is the
total energy flux density, ${\bf U}=\left(\frac{1}{2}\rho
u^{2}+\rho\Phi+\frac{\gamma}{\gamma-1}P_{th}\right){\bf u}+B^{2}{\bf
u}-u\cdot {\bf B}{\bf B}$ is the energy flux density and ${\bf Y}={\bf
u}{\bf B}-{\bf B}{\bf u}$. In these equations ${\bf B}$, ${\bf u}$,
$\Phi$ and $P_{th}$ represent the magnetic field vector, velocity
vector, Galactic gravitational field and the thermal pressure,
respectively. The latter is related to the gas internal energy
density ($e_{th}$) by $P_{th}=(\gamma-1)e_{th}$, with $\gamma=5/3$.
$S_{\rho}$, $S_{\Pi}$ and $S_{E}$ are the mass, momentum and energy
sources, respectively: $S_{\rho}=\frac{\Delta M_{SN}}{\Delta t}$
reproduces the injection of mass per unit volume, $\Delta M_{SN}$,
by supernovae at instant time $\Delta t$;
$S_{\Pi}=-\rho\nabla\Phi+\frac{\Delta \Pi_{SN}}{\Delta t}$ describes
the motion of the gas in the gravitational field and the momentum
injection per unit volume, $\Delta \Pi_{SN}$, by SNe at instant
$\Delta t$; and $S_{E}=\Gamma(z)-\Lambda+\frac{\Delta E_{SN}}{\Delta
t}$ describes the energy gains due to background heating,
$\Gamma(z)$, due to starlight varying with $z$ (Wolfire et al. 1995)
and SNe energy injection per unit volume, $\Delta E_{SN}$, at
instant time $\Delta t$ and the loss of energy by radiative cooling,
$\Lambda$, assuming that the gas is optically thin and in
collisional ionization equilibrium. The presence of background
heating leads here to the formation in the ISM of thermally stable
regimes at $\temp \leq 200$ K and $10^{3.9}< \temp\leq 10^{4.2}$ K.\\
\indent The model includes supernovae types Ia (randomly distributed in the
space and having a scale height of 325 pc) and II (randomly
distributed or clustered in OB associations formed in regions with
density and temperature thresholds of n$\geq 10$ cm$^{-3}$ and T$\leq
100$ K, respectively). The number and masses of the OB stars are
determined by an appropriate IMF following Massey et al. 1995; each
star explodes after its main sequence lifetime has elapsed, using a
time estimate by Stothers (1972). The trajectories of all O and B
stars are followed kinematically by attributing to each a random
velocity at time of formation so that the location of each exploding star is known at any time. For further details we refer to
Paper~I.
\begin{figure}[thbp]
\centering
\vspace*{4cm}
Jpeg image mavillez$\_$fig01.jpg
\vspace*{4cm}

\caption{Time evolution of the random field component (red), the mean field (green) and total field strength (black).
\label{fieldtot}
}
\end{figure}
The initial conditions comprise the setup of (i) the fixed vertical
gravitational field $\Phi$ provided by the stars in the disk, of
(ii) the interstellar gas having a density stratification
distribution that includes the cold, cool, warm, ionized and hot
``phases'' in the Galaxy, and (iii) a magnetic field with uniform
and random components given by $\bu=(\buz (n(z)/n_{0})^{1/2},0,0)$
and $\br=0$, respectively; here $\buz=3~\mu$G is the uniform
component field strength at the Galactic midplane, $n(z)$ is the
number density of the gas as a function of distance from the
Galactic midplane, where it has a value of $n_0=1\, {\rm cm}^{-3}$.
Periodic boundary conditions are applied along the boundaries
perpendicular to the Galactic plane (i.e., the $z$-direction), while
outflow boundary conditions are imposed at the top ($z=10$ kpc) and
bottom ($z=-10$ kpc) surfaces (parallel to the galactic plane). The
resolution in the grid varies on the fly according to density and/or
pressure gradients thresholds with 1.25 pc being the highest
resolution of three levels of refinements, achieved in the region
$\left|z\right|\leq 1$ kpc, while on the rest of the domain the
finest resolution is 2.5 pc (two levels of refinement). The
resolution of the coarse grid is 10 pc.

\begin{figure*}[thbp]
\centering

\vspace*{10cm}
Jpeg images mavillez$\_$fig02a.jpg and mavillez$\_$fig02b.jpg
\vspace*{10cm}
\caption{Two-dimensional slices through the three-dimensional data cube showing
the vertical (perpendicular to the midplane) distribution of the
density (left panel) and magnetic field (right panel) at time
$t=221.9$ Myr. The mean field in the midplane is $3.1~\mu$G. The $z$-scale above 0.5 and below -0.5 kpc is shrunk (in order to fit the paper size) and thus,
the distribution of the labels is not uniform. Red in the density map
refers to lowest density (and thus highest temperature if in pressure
equilibrium), while dark blue refers to the highest density (and thus
lowest temperature). In the magnetic field image red/blue correspond
to the highest/lowest strength. The left panel shows the presence of a
wiggly and thin disk of cold gas overlayed by a frothy thick disk
composed of neutral (light blue) and ionized (greenish) gas. The
expansion of superbubbles can be nicely seen, especially in the
southern hemisphere. The magnetic field map shows the thin magnetized
disk overlayed by loops of field lines, magnetic islands, and clouds
wrapped in field lines moving downwards. From these maps it can be
immediately seen that the simulations must take into account
large heights on either side of the midplane, allowing for the setup
of the disk-halo-disk circulation.
\label{dh}
}
\end{figure*}

\section{Results}

\subsection{Global evolution}

\begin{figure*}
\centering
\vspace*{7.6cm}
Jpeg images mavillez$\_$fig03a.jpg - mavillez$\_$fig03d.jpg
\vspace*{7.6cm}
\caption{Two-dimensional slices through the three-dimensional data
set at $z=0$ (Galactic midplane), showing density (top left panel),
temperature (top right panel), pressure $\pre/k$ (bottom left panel),
and magnetic field (bottom right panel) after 358 Myr of
evolution. Colour bars indicate the logarithmic scale of each
quantity. Red refers to lowest density, highest temperature and
pressure, while dark blue refers to the highest density and lowest
temperature and pressure. In the magnetic field image red/blue
correspond to the highest/lowest strength. Direct visual inspection
shows a strong inverse spatial correlation between gas density and
temperature (although the pressure diagram is still far from uniform),
whereas the correlation between magnetic field strength and gas
density is only present for the cold dense regions. For higher
temperature regions this correlation becomes rather poor and vanishes
completely for the hot gas (s.~Fig.~\ref{scatterDB}).
\label{fig1}
}
\end{figure*}

The initial evolution of the disk and halo gases is characterized by
the build-up of the random component of the magnetic field in addition
to the establishment of the duty disk-halo-disk cycle (see
Paper~I). The random field component is generated during the first 15
Myrs as a result of turbulent motions mainly induced by shear flows
due to SN explosions. For the remainder of the evolution time the
random field component and the mean field oscillate around their mean
values of 3.2 and $3.1~\mu$G (Figure~\ref{fieldtot}), respectively,
and the average total field strength amounts to $\btot = \sqrt{\bu^2 +
(\br)^2} \simeq 4.45~\mu$G. The build-up of the disk pressure by
supernovae and the establishment of the duty cycle, is similar to that
seen in the HD runs discussed in Paper I (see also Avillez 2000), that
is, the initially stratified distribution does not hold for long as a
result of the lack of equilibrium between gravity and (thermal,
kinetic and turbulent) pressure during the ``switch-on phase'' of SN
activity. As a consequence the gas in the upper and lower parts of the
grid collapses onto the midplane, leaving low density material behind.
However, in the MHD run it takes a longer time for the gas to descend
towards the disk. The complete collapse onto the midplane, like in the
hydrodynamical (HD) case, is prevented as a result of the opposing
magnetic pressure and tension forces. As soon as enough SNe have gone
off in the disk building up the required pressure support, transport
into the halo is not prevented, although the escape of the gas takes a
few tens of Myr to occur -- somewhat longer than in the pure HD
case. The crucial point is that a huge thermal overpressure due to
combined SN explosions can sweep the magnetic field into dense
filaments and punch holes into the extended warm and ionized H{\sc i}
layers as seen in Figure~\ref{dh}, which shows the density and
magnetic field distribution in a plane perpendicular to the Galactic
midplane at time $t=221.9$ Myr (the warm neutral and ionized H{\sc i}
gases are the light blue and greenish regions seen in the density
distribution). Once such pressure release valves have been set up,
there is no way from keeping the hot overpressured plasma to follow
the pressure gradient into the halo. As a consequence, transport into
the halo is not prevented and the duty disk-halo-disk cycle of the hot
gas is fully established, in which the competition of energy input and
losses into the ISM by SNe, diffuse heating, radiative cooling and
magnetic pressure leads the system to evolve into a dynamical
equilibrium state within a few hundred Myr (for an analytical estimate
see Paper~I). This time scale is considerably longer than that quoted
in other papers (e.g., Korpi et al. 1999, Kim et al. 2001), because
these authors do not take the galactic fountain into account. The
vertical structure of the thick disk and halo is addressed in more
detail in a forthcoming paper.

Figure~\ref{fig1} shows slices of the three-dimensional data cube of
the density, temperature, pressure $\pre/k$, and magnetic field
distributions in the Galactic midplane at 358 Myr of evolution with a
resolution of 1.25 pc. We can directly compare the morphology of the
ISM between this run and the similar HD run discussed in Paper~I. In
both cases the highest density (and lowest temperature, if the gas had
enough time to cool) gas tends to be confined into sheet-like
structures (filaments in 2D) which constitute shock compressed layers
(SCLs) as a result of the compression by shocks mainly driven by SNe
in their neighbourhood. The SCL structures are thicker in the MHD run
than in the HD run, because the compression of a shock wave at given
Mach number is less as energy has now to be shared between gas and
magnetic field. The $\temp< 10^{3}$ K gas in the MHD run tends to be
aligned with the magnetic field, while in the HD run there is no
preferred orientation. The highest densities up to 800 cm$^{-3}$,
observed in both runs occur in regions where several streams of
convergent flow meet. The orientation of these streams is random. We
note in passing, that if clouds are hit by SCLs from random
directions, turbulence in the interior is generated. Tapping the
turbulent ISM energy reservoir, which is huge, represents a neat way
of sustaining supersonic turbulence inside molecular clouds, which as
it is known, dissipates very efficiently. We will explore this idea in
more detail in a forthcoming paper.

In the MHD run cold denser regions are dominated by the magnetic field
and are more filamentary in structure than low density regions (which
have low field strengths, although, as seen in Fig.~\ref{scatterDB}, a
wide range of B-field strengths is spanned) due to the anisotropy
introduced by the strong field, which is much better correlated with
the gas density for lower than for higher temperature gas.

\subsection{Evolution of supernova remnants and superbubbles in
the magnetized ISM} \label{sbevol}
The evolution of magnetized SNRs and SBs has been extensively
studied in the literature (see e.g. Kulsrud et al. 1965, Summers
1975, Guiliani 1982). In particular the question of break-out has
been addressed in numerical studies by Tomisaka (1990) and Mineshige
et al. (1993), who both come to the conclusions that bubbles could
be confined to the gaseous disk. Since this is in disagreement with
our simulations we will discuss the subject of SB evolution (which
includes SNRs as a special case) in an ambient medium with frozen-in
magnetic field in more detail.

During their initial expansion phase, young supernova remnants are
roughly spherical as long as the radius is below or at the order of
the magnetic field curvature scale, but as time progresses magnetic
tension forces increase as the radius of curvature of the field
lines decreases as $F_{\rm t} \sim \bu^2/(4 \pi R_c)$, where $R_c$
and $F_{\rm t}$ are the radius of field curvature and tension force,
respectively, and $\bu$ is the uniform field component. It is
straightforward to estimate the average radius of a superbubble, or
respectively the time, when the shell stalls, because in the long
run magnetic tension forces beat the thermal pressure force in the
hot interior, since tension increases and pressure decreases with
time. This argument assumes some kind of ideal configuration, e.g.\
that the shock normal is perpendicular to the magnetic field
direction, which as can be seen in the simulations, does in general
not exist, as the field is tangled and inhomogeneous. It should also
be kept in mind that such an analytic analysis ignores the
possibility of MHD instabilities that would promote mixing with the
ambient medium and escape of hot bubble gas.

In the following we assume that the outer shell will have
cooled off and collapsed into a thin shell before slowing down
considerably, being subject to magnetic and thermal pressure. Let
$R_s$ be the shell radius, $\theta$ be the polar angle between the
radial and the magnetic field direction $\vec{B}$, and $\delta R$ the
shell thickness, so that the bubble radius is given by $R_b = R_s -
\delta R$. For $\theta = 0$ the outer shock is purely gasdynamic, as
compression along the B-field is unimpeded. However flux conservation
causes the field to wrap around the bubble and carry a tangential
component as $\theta$ increases. The bending of the lines of force
near the polar region leads to a deflection of the incoming flow and
thus to a $\theta$-dependent momentum transport. As this reduces the
inertia of the shell near the polar region, the deceleration there is
more effective leading to a ``dimple'' (s. Ferri\`ere et
al. 1991). Whereas the pressure is purely thermal here, the magnetic
pressure increases with $\theta$ and will dominate completely for
$\theta \rightarrow \pi/2$. Flux conservation at co-latitude $\theta$
in the thin shell approximation requires
\begin{eqnarray}
\int \vec{B} \vec{df} &=& \pi R_s^2 \sin^2\theta B_0 =
\int_{R_b}^{R_s} 2\pi R dR B_\theta \sin\theta \nonumber\\
&=& \pi (R_s^2 - R_b^2) B_\theta \sin\theta \,,
\end{eqnarray}
and since $R_b^2 = (R_s - \delta R)^2 \approx R_s^2 - 2 R_s \delta
R$, we obtain
\begin{equation}
B_\theta \approx \frac{1}{2} B_0 \frac{R_s}{\delta R} \sin\theta
\,, \label{btheta}
\end{equation}
which at the equatorial region, where field tension is strongest,
yields $B_\theta = \frac{1}{2} B_0 \frac{R_s}{\delta R}$, where
$B_\theta$ is the poloidal field component in the shell.

Numerical simulations show (e.g.\ Ferri\`ere et al. 1991) that in
case of interstellar magnetic field strengths of the order $3-5 \,
\mu{\rm G}$ the evolution until the outer shock decays into a fast
magnetosonic wave can still be well described by a similarity
solution (assuming a homogeneous ambient medium) given by McCray \&
Kafatos (1987). In the case of quasi-sphericity, the momentum
equation for self-similar flow, and using Eq.~(\ref{btheta}), reads
\begin{eqnarray}
\frac{d}{dt} (M_s \dot R_s) &=& 4\pi \left(R_s^2 P_b -
\frac{B_\theta^2}{4\pi} R_s \delta R \right) \nonumber\\
&=& 4\pi R_s^2 \left(P_b - P_{m0} \frac{R_s}{2 \delta R} \right)\,.
\label{dyneq1}
\end{eqnarray}
Here $P_b$ and $P_{m0}=\frac{B_0^2}{8\pi}$ are the thermal pressure
in the bubble and the magnetic pressure in the ambient medium,
respectively. The bubble expansion will start to reverse, once the
force acting on the shell points inwards. In other words,
the radius or time of maximum extension of the bubble in the
equatorial direction can be roughly estimated by demanding that the
term in brackets on the right-hand side of Eq.~(\ref{dyneq1})
vanishes, giving
\begin{equation}
P_b = P_{m0} \frac{R_s}{2 \delta R} \,.
\label{maxexp}
\end{equation}

The applicability of similarity solutions (cf.\ Weaver et al. 1977)
provides $P_b$ and $R_s$ by
\begin{eqnarray}
  R_s &=& A \,\left(\frac{L_*}{\rho_0}\right)^{1/5} \, t^{3/5} \\
  \label{simsol1}
  A &=& \left(\frac{125}{154 \pi}\right)^{1/5} \, \nonumber \\
  P_b &=& D \,L_*^{2/5} \, \rho_0^{3/5} \, t^{-4/5} \\
  D &=& \frac{7}{(3850 \pi)^{2/5}} \nonumber \,,
  \label{simsol2}
\end{eqnarray}
with $L_*$ denoting the mechanical luminosity generated by stellar
wind and supernova activity of the central star cluster, and
$\rho_0=const.$ is the ambient mass density. Since SNe are the
dominant source for rich clusters one can use for a simple estimate,
$L_* = N_* E_{SN}/\tau_{\rm MS} = 3.17 \, N_* E_{51}/\tau_7 \,
\times 10^{36} \, {\rm erg} \, {\rm s}^{-1}$, where $N_*$ is the
number of supernovae, $E_{SN}$ and $E_{51}$,
$\tau_{\rm MS}$ and $t_7$ are the supernova energy, also measured in
$10^{51}$ ergs, respectively, and the main sequence life time of the
least massive star, also in units of 10 Myrs, respectively.

To solve Eq.~(\ref{maxexp}) we need an expression for the relative
shell thickness, $\delta R$. In the equatorial region, the pressure
in the shell at later times will be dominated by magnetic pressure
so that momentum conservation across the shock requires
\begin{equation}
\rho_0 \dot R_s^2 \simeq \frac{B_\theta^2}{8 \pi} \,,
\label{momcon1}
\end{equation}
and using Eq.~(\ref{btheta})
\begin{eqnarray}
\dot R_s^2 &=& \frac{B_0^2}{32 \pi \rho_0} \, \left(
\frac{R_s}{\delta
R}\right)^2 \sin^2\theta \nonumber\\
\Rightarrow \frac{\delta R}{R_s} &=& \frac{1}{2\sqrt{2}}
\frac{V_{\rm A0}}{\dot R_s} \sin\theta \,,
\label{momcon2}
\end{eqnarray}
implicitly assuming that the shock propagates in an undisturbed
ambient medium, with $V_{\rm A0}=\frac{B_0}{\sqrt{4\pi \rho_0}}$
being the Alfv\'en speed there. Eq.~(\ref{momcon2}) shows the
thickening of the shell with time as it decelerates, like $\delta R
\propto R_s/{\dot R_s} \propto t$:
\begin{eqnarray}
\delta R &=& \frac{5}{6\sqrt{2}} V_{\rm A0} \, t \sin\theta \\
\Leftrightarrow \delta R &=& 1.2 \, \frac{B_{-6} \, t_6}
{\sqrt{n_0}}  \sin\theta \, {\rm pc}\,,
\label{shth1}
\end{eqnarray}
where $n_0$, $B_{-6}$ and $t_6$ denote the ambient number density in
units of ${\rm cm}^{-3}$, the ambient magnetic field strength in
units of $\mu$G, and the time in Myrs, respectively.

As long as the ambient thermal pressure is small compared to the
pressure inside the bubble and the magnetic pressure in the shell,
and combining Eqs.~(\ref{simsol1}), (\ref{simsol2}),
(\ref{momcon2}), and inserting them into Eq.~(\ref{maxexp}), we
obtain for the time $t_{\rm max}$ of maximum bubble expansion (for
$\theta = \pi/2$, where tension forces are strongest)
\begin{eqnarray}
  t_{\rm max} &=& \kappa \left(\frac{L_* \rho_0^{3/2}}{B_0^5}
  \right)^{1/2} \\
  \kappa &=& \left(\frac{5}{3} \frac{D}{A}\right)^{5/2} \,
  (8 \pi)^{5/4}\\
   &=& 4.255 \\
  \Rightarrow t_{\rm max}  &=& 718.39 \, L_{37}^{1/2} \, n_0^{3/4} \,
  B_{-6}^{-5/2} \, {\rm Myr}
  \,;
  \label{tmax}
\end{eqnarray}
here $L_{37}$ denotes the mechanical luminosity in units of
$10^{37}$ erg/s. Note, that this is also roughly the time scale on
which the outer shock decays into a fast magnetosonic wave, which is
disturbing the ambient medium with a characteristic speed of
$a_f=\sqrt{\frac{k_B T_0}{\mu \, \bar m} + \frac{B_0^2}{4 \pi
\rho_0}}$, with $k_B$, $\mu$, $\bar m$ and $T_0$ denoting
Boltzmann's constant, the mean molecular weight, the mean particle
mass of the gas and the ambient temperature, respectively. Equating
this to the shock speed travelling outwards at $\dot R_s$, results
in a typical decay time of $t_f \simeq 285 \, L_{37}^{1/2} \,
n_0^{3/4} \, B_{-6}^{-5/2} \, \left(0.58 + \frac{T_3 n_0}{\mu
B_{-6}^2} \right)^{-5/4} {\rm Myr}$. Here $T_3$ denotes the ambient
temperature in units of $10^3 \, {\rm K}$. For example, taking a
field strength of $B_{-6}=5$, an ambient density of $n_0= 0.3$ and
$L_{37} = 3$, corresponding to a cluster with about 50 SNe, we
obtain $t_f \simeq 6.2 \, {\rm Myr}$. Since the dependence on $B_0$,
$n_0$, and $L_*$ is exactly the same as in Eq.~(\ref{tmax}), the
outer shock in the equatorial region will always have decayed just
about before the bubble will stall. Although several of the
assumptions made above will break down, the qualitative picture will
be approximately correct. In particular, the shell will be still
small compared to the bubble radius, i.e.\ $\frac{\delta R}{R_s} =
\frac{1}{2\sqrt{2}}\frac{V_{A0}}{a_f}=\frac{1}{2} (2+\beta)^{-1/2}$,
where $\beta = 8\pi n_0 k_B T_0/B_0^2$ is the plasma beta in the
ambient medium. Thus even for $\beta \rightarrow 0$, $\delta R \leq
0.35 R_s$.

The radius of maximum expansion (perpendicular to the B-field
direction) is therefore given by
\begin{eqnarray}
  \Rightarrow R_{\rm max}  &=& \left(\frac{125 \kappa^3}{154 \pi}
  \right)^{1/5} \, L_*^{1/2} \, \rho_0^{1/4} \, B_0^{-3/2} \\
   &=&  2217 \, L_{37}^{1/2} \, n_0^{1/4} \,
  B_{-6}^{-3/2} \, {\rm pc} \,.
  \label{rmax}
\end{eqnarray}
The maximum extension therefore depends most sensitively on the
magnetic field strength, and only weakly on the ambient density.
Using the above values for $B_0$, $n_0$ and $L_{37}$, we obtain
$R_{\rm max} \approx 254$ pc and $t_{\rm max} \approx 9$ Myr, well within our grid size and simulation timescale, respectively. With
density inhomogeneities and stratification perpendicular to the disk
left aside, the bubble is still not too aspherical. In the following
however, the contraction in the equatorial region will make it more
elongated towards the poles, i.e.\ along the magnetic field. In
addition, contraction will increase the pressure in the interior
thus acting against the tension forces, but also \textit{enhance
expansion predominantly along the field.} We therefore conclude that
elongated features in our simulations correspond to old
superbubbles, unless large density gradients in the ambient medium
have been present before. If the latter is the case, then the
bubbles continue early on to expand into the neighbouring regions
where the field lines have been moved aside by other bubbles. In any
case, the evolved bubbles lose their spherical form acquiring an
oval-like appearance (see Figure~\ref{fig1}). After 358 Myr of
evolution there are still a large number of regions were the mean
field maintains its initial orientation, although locally it is
heavily distorted by SNe (Figure~\ref{fig1}). Interestingly, the
average $P/k$ value is higher in the MHD case due to the magnetic
field contribution, which in the absence of dissipative processes
and loss of flux always is present.

Perpendicular to the galactic plane the expansion of bubbles will be
alleviated due to the density stratification of the extended disk
and Rayleigh-Taylor instabilities can occur once the shell
accelerates over a few density scale heights. The effect of the
magnetic field then will be similar to surface tension in an
ordinary fluid, limiting the growth of the most unstable modes to a
maximum wave number. It is therefore expected that shell fragments
or cloudlets should have a minimum size. A number of shells will
stall before that happens, and these will appear in simulations and
observations, like e.g.\ \object{NGC 4631} (Wang et al. 2001) as
extended loops.

\subsection{Field dependence with density}
\label{cslaw}

During the evolution of the system, thermal and dynamical processes
broaden the distribution of the field strength with density so that
after the global dynamical equilibrium has been set up, the field
strength in the disk spans two orders of magnitude from $0.1$ to $15$
$\mu$G (see Fig.~\ref{scatterDB}). The spreading in the field strength
increases with temperature, being largest for the hot ($\temp >
10^{5.5}$ K) and smallest (between 2 and 6 $\mu$G) for the cold
($\temp \leq 200$ K) gas. The lowest (0.1 $\mu$G) and higher (larger
than 10 $\mu$G) field strengths are associated with $\temp > 10^{4.2}$
gas and densities of 0.1-1 cm$^{-3}$.  The intermediate temperature
regimes have the bulk of their scatter points similarly distributed
between 2 and 6 $\mu$G, although there is a clear distinction between
the different temperature regimes with respect to the density
coverage. It can be seen that the $10^{4.2}<\temp\leq 10^{5.5}$ K gas
covers a large fraction of the scatter points overlapping with the
coolest and partially with the hot regimes, and there, mainly with
those of low densities and high field strengths. It almost completely
covers points located in the $10^{3.9}<\temp \leq 10^{4.2}$ K both in
the field strength and in the density distribution. The thermally
unstable gas ($200 <\temp \leq 10^{3.9}$ K) completely overlaps the
$\temp \leq 200$ K scatter points, although the latter covers smaller
field strength intervals than the unstable gas. The $10^{4.2}<\temp
\leq 10^{5.5}$ gas coverage with other (stable) regimes is an
indication that it is in a transient state trying to move towards
thermally stable regions at $\temp \leq 10^{4.2}$ K or to the
$\temp>10^{5.5}$ K range. Whether this gas becomes part of a
classically stable region, however, depends if it has enough time to
cool down (or to be heated up) with respect to dynamical time scales,
which change its location on the $p-V$-diagram. The higher field
strengths at the intermediate temperature regimes are related to the
increase in compression of the field lines in newly formed shells and
SCLs.

\begin{figure}[thbp]
\centering
\vspace*{4cm}
Jpeg image mavillez$\_$fig04.jpg
\vspace*{4cm}

\caption{Scatter plot of B versus $\rho$ for $\temp\leq 200$
  (black), $200<\temp\leq 10^{3.9}$ (orange), $10^{3.9}<\temp\leq
  10^{4.2}$ (blue), $10^{4.2}<\temp\leq10^{5.5}$ K (green), and
  $\temp>10^{5.5}$ K (red) regimes at 300 Myr of disk evolution. The
  points in the plot are sampled at intervals of four points in each
  direction.  Note that during the evolution of the system the field
  strength broadened its distribution spanning two orders of magnitude
  after 300 Myr.
  \label{scatterDB} }
\end{figure}
The large scatter in the field strength for the \textit{same} density,
seen in Figure~\ref{scatterDB}, suggests that the field is being
driven by the inertial motions, rather than it being the agent
determining the motions. In the latter case the field would not be
strongly distorted, and it would direct the motions predominantly
along the field lines. In ideal MHD field diffusion is negligible, and
the coupling between matter and field should be perfect (we are
of course aware of numerical diffusion, which weakens the argument at
sufficiently small scales). Therefore gas
compression is correlated with field compression, except for strictly
parallel flow. The high field variability is also seen in the right
bottom panel of Fig.~\ref{fig1}, which shows a highly turbulent field,
that seems to be uncorrelated with the density, and thus, the
classical scaling law $B\sim\rho^\alpha$, with $\alpha=1/2$, according
to the Chandrasekhar-Fermi (CF) model (1953) will not hold. It should
be kept in mind that in CF it was assumed that the field is distorted
by turbulent motions that were subalfv\'enic, whereas in our
simulations in addition both supersonic and superalfv\'enic motions
can occur, leading to strong MHD shocks. In other words, according to
the CF model, the turbulent velocity is directly proportional to the
Alfv\'en speed, which in a SN driven ISM need not be the case.
If ram pressure fluctuations in the ISM
are dominant as our simulations suggest, we would indeed not expect a
perfect correlation with $\alpha=1/2$ but a broad distribution of $B$
versus $\rho$. Although in general $0\leq \alpha \leq 1$ would be
expected, it should be noted that in reality heating and cooling
processes, and even magnetic reconnection could induce further
changes, making the correlation rather complex.

\subsection{Weight of thermal, magnetic and ram pressures}

\begin{figure}[thbp]
\centering
\vspace*{4cm}
Jpeg image mavillez$\_$fig05.jpg
\vspace*{4cm}
\caption{
Comparison of the averages $\langle$P$_{ram}\rangle$ (green),
$\langle$P$_{th}\rangle$ (black), and $\langle$P$_{mag}\rangle$ (red)
as function of the temperature (in the simulated disk
$\left|z\right|\leq 250$ pc) averaged over temperature bins of $\Delta
\log \temp=0.1$ K.
\label{rampressure2}
}
\end{figure}

In order to gain insight into the driving forces in the simulated disk
($\left|z\right|\leq 250$ pc) we turn now to the relative weight of
the ram, thermal and magnetic pressures as function of the temperature
averaged over bins of $\Delta \log \temp=0.1$ K
(Fig.~\ref{rampressure2}). The average distributions shown in the
figure have been calculated by using 51 snapshots with a time interval
of 1 Myr during the time evolution of 350 and 400 Myr.

The relative importance of thermal pressure over ram and magnetic
pressures increases with temperature. The gas with $\temp \geq 10^{5.5}$ K is mainly thermally driven, while at lower temperatures thermal
pressure has no influence on the dynamics of the flow. For $\temp \leq
150$ K, $\pre_{th}<\pre_{ram}<\pre_{mag}$ indicating that the gas is
dominated by the Lorentz $\vec{j} \times \vec{{\rm B}}$ forces, and
the magnetic field determines the motion of the fluid.  At the
intermediate temperatures ram pressure dominates, and therefore, the
magnetic pressure does not act as a significant restoring force (see
Passot \& V\'azquez-Semadeni 2003) as it was already suggested by the
lack of correlation between the field strength and the density. It is
also noteworthy that in this temperature range ($150 - 7.4\times 10^{5}$ K)
the weighted magnetic pressure is roughly constant, suggesting that
the magnetic and thermal pressures are largely independent, whereas
both thermal and ram pressures undergo large variations in this
temperature interval.

We therefore conclude that in galactic disk environments the magnetic
field will have a strong influence on the cold gas,
particularly by controlling cloud and star formation processes. On the
other hand, the warm and hot phases of the ISM, which are dominating
by volume but not by mass, are affected \textit{dynamically } only to
a minor extent by magnetic pressure forces. However, it should be kept
in mind, that microscopic processes, like heat conduction, will be
strongly modified by the presence of a magnetic field.
\begin{figure}[!thbp]
\centering
\vspace*{11cm}

Jpeg images mavillez$\_$fig06a-c.jpg

\vspace*{11cm}

\caption{Averaged volume weighted histograms of the thermal (top
panel), magnetic (middle panel) and total (bottom panel) pressures
for the different temperature regimes in the ISM. The total
distribution of each pressure is shown by the dashed black line. The
histograms have been calculated using 51 snapshots with a time
interval of 1 Myr between 350 and 400 Myr. \label{mhdpdfs} }
\end{figure}

\subsection{Thermal and magnetic pressures}

The averaged (over the period 350-400 Myr) volume weighted histograms
of the thermal, magnetic and total ($\pre_{tot}=\pre_{th}+\pre_{mag}$,
which is calculated at each grid cell, and then used in the
construction of the corresponding histogram) pressures of the
different temperature regimes in the magnetized ISM
(see Fig.~\ref{mhdpdfs}) show that for volume fractions dN/N$\geq 10^{-4}$
the full distributions (shown by the dashed lines) have a pressure
coverage running from three (for $\pre_{tot}$) to more than five (for
$\pre_{th}$) orders of magnitude and having smooth (for $\pre_{th}$)
and steep (for $\pre_{mag}$ and $\pre_{tot}$) peaks located in the
$[2\times 10^{-13}, 2\times 10^{-12}]$ dyne cm$^{-2}$ interval.  These
peaks are determined by the contributions of the $200<\temp\leq
10^{3.9}$ K and $10^{4.2}<\temp\leq 10^{5.5}$ K regimes followed by
the hot gas. The thermally stable regimes $10^{3.9}<\temp\leq
10^{4.2}$ K and $\temp
\leq 200$ K, by this order, have the smallest contribution to the
peaks.

In the thermal and magnetic pressure histograms it is possible to
distinguish two regions ($\pre_{th} <10^{-13}$ and $\pre_{th}
>10^{-12}$ dyne cm$^{-2}$), which are determined by the gas with $\temp
\leq 10^{3.9}$ and $\temp > 10^{4.2}$ K (with the hot gas being the
main contributor for the high pressure region), respectively. In the
$\pre_{mag}$ histogram (that has a conical structure covering the
$[10^{-15}, 5\times 10^{-12}]$ dyne cm$^{-2}$ region for dN/N $\geq
10^{-4}$) all regimes contribute to these two regions, although the
hot gas is the main contributor to the low pressure region and the
cold gas contributes mainly to the high pressure range.

A comparison between the top and middle panels of the figure indicates
that the highest values of the total pressure, i.e., $\pre_{tot} \geq
10^{-12}$ dyne cm$^{-2}$, are mainly of thermal origin. Contributions
come from the gas with $\temp >10^{4.2}$ K and in particular from the
hot gas (whose histogram has a central peak located around $2\times
10^{-12}$ dyne cm$^{-2}$ and two smaller peaks in the right wing of
the histogram at $\pre_{th}>10^{-11}$ dyne cm$^{-2}$, resulting from
SN events). The gas with $200 <\temp\leq 10^{4.2}$ has a very small
contribution and the cold gas practically none. For the remaining
pressure range ($\pre_{tot} < 10^{-12}$ dyne cm$^{-2}$) both magnetic
and thermal pressures contribute, with the magnetic contributions
having a larger weight than the thermal ones.

\subsection{Volume weighted histograms}

In order to gain insight into the distribution of the disk gas as a
function of density or temperature, it is useful to construct volume
weighted histograms. They give us information about the occupation
fraction of a certain ``gas phase'' with respect to volume averaged
density and temperature. Therefore a peak in the histogram tells us,
at which density/temperature a particular phase is concentrated,
because here the occupation fraction is highest.
A comparison between the averaged (over the time 350 through 400
Myr) volume weighted histograms of the density and temperature of
the magnetized and unmagnetized disk gas (Fig.~\ref{pdfd}) shows
noticeable differences that include the (i) decrease/increase by an
almost order of magnitude in the histograms density/temperature
coverage, (ii) change in the relative weight of the dominant
temperature regimes (in the density histograms) and consequently
(iii) changes in the pronounced bimodality of the total density and
temperature histograms.
\begin{figure*}[thbp]
\centering
\vspace*{7.6cm}
Jpeg images mavillez$\_$fig07a.jpg - mavillez$\_$fig07d.jpg
\vspace*{7.6cm}

\caption{Averaged volume weighted histograms of the density for the
different temperature regimes of the disk gas (the full distribution
is shown by the dashed line) in the top panels and averaged
temperature histograms in the bottom panels for the MHD run (left
column) and the HD run (right column). The histograms have been
calculated using 51 snapshots with a time interval of 1 Myr between
350 and 400 Myr.
\label{pdfd} }
\end{figure*}

The density histograms in both HD and MHD runs are dominated by the
$\temp > 10^{5.5}$ K (less important), $10^{4.2}<\temp\leq 10^{5.5}$
K and $200<\temp\leq 10^{3.9}$ K regimes. The latter regime is the
most important in the HD run contributing to the pronounced peak
around $\mbox{n}\sim 1$ cm$^{-3}$ in the total density distribution,
whereas in the magnetized ISM the latter two regimes interchange in
importance (there is a strong reduction in the peak of the
$200<\temp\leq 10^{3.9}$ K gas). This leads to a shift of the peak in
the total density histogram to $\mbox{n}\sim 0.2$ cm$^{-3}$, where it
is mainly dominated by the $10^{4.2}<\temp\leq 10^{5.5}$ K gas
contribution. The hot gas averaged PDFs have a similar shape but
different density coverage, being larger in the HD run by an order
of magnitude and peak location (being a factor of two higher in the
MHD run). This suggests that the areas underneath the histograms, that
is, the volume filling factors of the hot gas, are not much
different in the two runs. The smallest peaks seen in the density
histograms belong to the thermally stable regimes $\temp \leq 200$ K
(which has the smallest occupation fraction), followed by the
$10^{3.9}<\temp\leq 10^{4.2}$ K regime. With the inclusion of the
magnetic field, there is an increase in the density coverage by the
cold gas histogram and thus of its occupation fraction in the disk.

The variations seen in the density weighted histograms are also seen
in temperature PDFs having different structures. In effect, while the
temperature PDF of the HD run has a bimodal structure (as it has two
peaks: one at 2000 K and another around $10^{6}$ K), in the MHD run
the decrease/increase in importance of the $10^{3.9} <\temp
\leq 10^{4.2}$ K/$10^{4.2} <\temp \leq 10^{5.5}$ K leads to the
reduction/increase of the occupation fraction of these regimes and
therefore to a change of the histogram structure appearing it to be
unimodal. This variation of the intermediate region appears to be an
effect of the presence of the magnetic field, with the smoothing
effect being less pronounced for lower field strengths in the disk.

\begin{figure}[thbp]
\centering
\vspace*{6cm}
Jpeg images mavillez$\_$fig08a-b.jpg

\vspace*{6cm}

\caption{
Time history of the volume occupation fractions of the different
temperature regimes for the MHD (top panel) and HD (bottom panel)
runs.
\label{vff}}
\end{figure}

\subsection{Volume and mass fractions}

During most of the history ($t> 100$ Myr) of ISM simulation the
occupation ($\mbox{f}_{\mbox{v}}$) and mass
($\mbox{f}_{\mbox{{\small M}}}$) fractions of the different thermal
regimes (Figs.~\ref{vff} and \ref{figmass} -- top and middle panels)
have an almost constant distribution, varying around their mean
values (cf.~Table 1). The thermally stable regimes with $\temp\leq
200$ K and $10^{3.9}<\temp \leq 10^{4.2}$ K have similar occupation
fractions of $\sim 5\%$ and $\sim 10\%$, respectively, in both runs,
while the hot gas has an increase from $\sim 17\%$ in the HD run to
$\sim 20\%$ in the MHD case. By far most of the disk volume is occupied by
gas in the thermally unstable regimes at $200<\temp \leq 10^{3.9}$
and $10^{4.2}<\temp \leq 10^{5.5}$ K with similar occupation
fractions $\sim 30\%$ in the MHD run, while in the HD run these
regimes occupy $46\%$ and $22\%$, respectively, of the disk volume.

The major fraction of the disk \emph{mass} ($\sim 93\%$ and $\sim
80\%$ in the HD and MHD runs, respectively) is occupied by the gas
with $\temp \leq 10^{3.9}$ and transferred between the cold and the
thermally unstable regimes with the corresponding mass changes. The
cold ($\temp\leq 200$ K) regime during most of the simulation time
encloses more mass than in the MHD run until $\sim 320$ Myr when
both runs show a similar amount of cold mass ($\sim 40\%$), while
the thermally unstable regime encloses the remaining mass. This
results from the fact that in the MHD run the gas compression is
smaller than in the HD case, because part of the shock energy goes
into work against magnetic pressure. The remaining ISM mass is
distributed between the other temperature regimes with the $10^{3.9}
<\temp\leq 10^{4.2}$ K and $10^{4.2} <\temp\leq 10^{5.5}$ K regimes
enclosing a total of $\sim 7\%$ and $\sim 16\%$ of mass in the HD
and MHD runs, respectively, and the hot gas enclosing $<1\%$ of the
disk mass in both runs.

For most of the simulation time, in both runs, 60-70\% of the warm
neutral mass ($500<\temp \leq 8000$ K) is contained in the
$500\leq\temp\leq 5000$ K temperature range (bottom panel of
Figure~\ref{figmass}). This result in excellent agreement with the
observations of Fitzpatrick \& Spitzer (1997) who found a value of
$\approx 60$\% and Heiles \& Troland (2003) who determined a lower
limit of $\approx 50$\% (see also Heiles 2001).

\begin{figure}[thbp]
\centering

\vspace*{11cm}

Jpeg images mavillez$\_$fig09a-c.jpg

\vspace*{11cm}

\caption{Top and middle panels show the histories of the mass
fractions of the different thermal regimes for the magnetized (top)
and unmagnetized (middle) ISM. Bottom panel shows the history of the
fraction of mass of the WNM gas having $500 <\temp \leq 5000$ K in
the disk for the HD (red) and MHD (black) runs. \label{figmass} }
\end{figure}

\subsection{Turbulent velocities}

In both the MHD and HD runs the root mean square (rms) velocity,
$\vrms$, which is a measure of the disordered motion of the gas,
increases with temperature as can be seen in Figure~\ref{vrms},
which shows the $\vrms$ for the different thermal regimes in the MHD
(top panel) and HD (bottom panel) runs. The average rms velocity
($\langle\vrms\rangle$) in the last 100 Myr of evolution of both
media for the different thermal regimes is shown in Table 1. The
figure and table show that there are large fluctuations in the
$\vrms$ of the different thermal regimes in the HD run, which are
reduced due to the presence of the magnetic field.

The rms velocities are in good agreement with those
estimated from observations of cold/cool neutral
($7-10$ km/s; e.g., Kulkarni \& Fich 1985; Kulkarni \& Heiles 1987), warm neutral ($\sim 14$ km/s; Kulkarni \& Fich 1985), warm ionized ($\leq 30$ km/s;
Reynolds 1985) and hot ($\sim 33-78$ km/s; e.g., Zsarg\'o et al. 2003) gases.
Again, the near constancy of the rms velocity with time indicates the
presence of a dynamical equilibrium, with random motions, i.e.\
thermal and turbulent pressures adding to the total pressures,
provided that the energy injection rate remains constant on a global
scale.

\begin{table}
\centering \caption{ Summary of the average values of volume filling
factors, mass fractions and root mean square velocities of the disk
gas at the different thermal regimes for the HD and MHD
runs\label{table1}}
\begin{tabular}{c|cc|cc|cc}
\hline
\hline
T  & \multicolumn{2}{c|}{$\langle \mbox{f}_{\mbox{v}}\rangle $$^{a}$ [\%]}&
\multicolumn{2}{c|}{$\langle \mbox{f}_{\mbox{M}}\rangle $$^{b}$ [\%]} &
\multicolumn{2}{c}{$\langle \mbox{v}_{\mbox{rms}}\rangle$$^{c}$}\\
\cline{2-7}
[K] & HD & MHD & HD & MHD & HD & MHD\\
\hline
$<200$ K              & ~5 & ~6 &  44.2 & 39.9 & ~7 & 10 \\
$200-10^{3.9}$        & 46 & 29 &  49.0 & 43.7 & 15 & 15 \\
$10^{3.9}-10^{4.2}$   & 10 & 11 &  ~4.4 & ~8.5 & 25 & 21 \\
$10^{4.2}-10^{5.5}$   & 22 & 33 &  ~2.0 & ~7.4 & 39 & 28 \\
$>10^{5.5}$           & 17 & 21 &  ~0.3 & ~0.5 & 70 & 55 \\
\hline
\multicolumn{7}{l}{$^a$ Occupation fraction.}\\
\multicolumn{7}{l}{$^b$ Mass fraction.}\\
\multicolumn{7}{l}{$^c$ Root mean square velocity in units of km s$^{-1}$.}
\end{tabular}
\end{table}

\begin{figure}[thbp]
\centering
\vspace*{6cm}
Jpeg images mavillez$\_$fig10a-b.jpg

\vspace*{6cm}
\caption{
History of the root mean square velocity of the different temperature
regimes of the disk gas in the MHD (top panel) and HD (bottom panel)
runs.
\label{vrms} }
\end{figure}

\section{Discussion}
\label{disc}

In the previous sections we compared high resolution
three-dimensional kpc-scale HD and MHD simulations of the ISM,
driven by SNe at the Galactic rate, fully tracking the
time-dependent evolution of the large scale Galactic fountain (up to
10 kpc on either side of the Galactic midplane) for a time
sufficiently long (400 Myr) so that the memory of the initial
conditions is completely erased, and a global dynamical equilibrium
is established. These simulations show how important the
establishment of the duty cycle is, and how it affects the global
properties of the ISM, e.g., occupation fraction of the hot gas,
just to name one. An essential feature of the hot gas is its escape
into the halo establishing the fountain flow, thereby reducing the
expansion volume in the underlying disk significantly. It is not
possible to suppress such a flow even in the presence of an
obstructing disk parallel magnetic field. All that is needed for
break-out of the gas into the halo is a sufficient overpressure in
the superbubbles with respect to the ambient medium. The effect of
the magnetic field is just adding another pressure component.
Its topology is almost irrelevant unless the galactic magnetic field
is extremely high and/or the SN rate is very low.
The field may inhibit the break-out of an individual remnant, but
certainly not the high-pressure flow resulting from supernova
explosions in concert within an OB association. Thus, the occupation
fractions of the hot gas in the disk in the HD and MHD simulations are
not too much different, although there is a slight increase in the MHD
run as a result of the magnetic tension forces aiding to confine
bubbles and Lorentz forces obstructing mixing with cooler gas. In
Paper~I we have shown that the volume filling factor of the hot regime
strongly correlates with the supernova (and therefore star formation)
rate. Although we have not carried out MHD simulations for higher
supernova rates, it seems very plausible that this correlation will
persist in the MHD case, since higher rates will produce more hot
plasma which, as we have shown here, is not magnetically controlled.

We also emphasize the importance of carrying out
\emph{three-dimensional} simulations in order to describe the
evolution of a magnetized ISM adequately. The idea that a disk
parallel magnetic field could suppress break-out and outflow into the
halo was mainly based on 2D simulations carried out in the last 15
years, owing to computing power limitations. It is obvious, that in
2D-MHD, the flow perpendicular to the magnetic field lines (and hence
to the galactic plane) is subject to opposing and ever increasing
magnetic tension and pressure forces. In 3D however, field lines can
be pushed aside and holes and channels can be punched into the
decreasing gas and field with $z$-height, allowing pressurized flow to
circumvent \textit{increasing tension and pressure forces} in
$z$-direction. It is exactly this behaviour that we see in our
simulations. This is less surprising if we think of similar problems
of terrestrial plasma confinement, although here plasma instabilities prevail.

The lack of correlation between the field strength and density found
in the present simulations is supported observationally from
measurements of the magnetic field strength in the cold neutral medium
(Troland \& Heiles 2001). A similar result is shown in the simulation
described in Kim et al. (2001), although these authors claim that the
magnetic field scales as $\rho^{0.4}$ at densities $n>1$ cm$^{-3}$.
Their Figure~2 shows an almost order of magnitude variation of the
field with the density for the coolest gas. However, the distribution
of the scatter points is somewhat different than that seen in
Figure~\ref{scatterDB} of the present paper. This discrepancy can be
explained by the (200~ pc)$^{3}$ box that Kim et al. have used,
centered in the Galactic midplane having no density stratification and
with periodic boundary conditions in all the box faces, driven by SNe
at a rate of 12 times the Galactic value and using an uniform field
strength of 5.8 $\mu$G orientated along the $x-$direction; thus they
were unable to describe the disk-halo-disk circulation and did not
allow for a global dynamical equilibrium to be established (see also
Mac Low et al. 2004).

The distribution of the disk mass in the warm neutral medium (WNM),
in particular in the WNM fraction within the classical thermally
unstable regime seen in the simulations is strongly supported by
interferometric (Kalberla et al. 1985). Moreover optical/UV
absorption-line measurements (Spitzer \& Fitzpatrick 1995,
Fitzpatrick \& Spitzer 1997) indicate that a large fraction ($\sim
63\%$) of the warm neutral medium (WNM) is in the unstable range
$500<\temp<5000$ K, whereas 21~cm line observations (Heiles 2001;
Heiles \& Troland 2003) provide a lower limit of $48\%$ for the WNM
gas in this unstable regime. Direct numerical simulations of the
nonlinear development of the thermal instability under ISM
conditions with radiative cooling and background heating discussed
in Gazol et al. (2001) and Kritsuk \& Norman (2002) show that about
60\% of the system mass is in the thermally unstable regime.
However, it is unclear from their simulations what is the time
evolution of this mass fraction and what is explicitly the origin of
the unstable gas, although these authors suggest that ensuing
turbulence is capable of replenishing gas in the thermally unstable
regime by constantly stirring up the ISM.  We have carried out
detailed numerical studies of the stability of the ISM gas phases
(Avillez \& Breitschwerdt 2005), and could verify the hypothesis
that SN driven turbulence is capable of replenishing fast cooling
gas in classically unstable regimes.

Nonetheless, it is interesting to ask why such a large amount of gas can
exist in a thermally unstable regime.  Obviously the simple Field
(1965) criterion, according to which instability is expected to set
in, viz.\ $\left(\frac{\partial \mathcal{L}}{\partial T} \right)_P
<0$, where $\mathcal{L}$ is the heat loss function per unit mass,
and T and P are the temperature and thermal pressure of the fluid
element, respectively, is not adequate. It is the turbulence that
can have a stabilizing effect thereby inhibiting local condensation
modes.  The situation is reminiscent to the existence of the solar
chromosphere, which consists of gas at around $10^5$ K, again in the
thermally unstable regime. The reason is because heat conduction can
prevent thermal runaway on small scales. In other words, diffusion
processes have a stabilizing effect. In our case it is turbulent
diffusion that replaces the r\^ole of conduction, again, most
efficient for large wavenumbers. The turbulent viscosity $\nu_{\rm
turb} \sim Re \, \nu_{\rm mol}$ can be orders of magnitude above the
molecular viscosity, with $Re$ being the Reynolds number of the
flow.  What happens physically then, is that with increasing eddy
wavenumber $k=2\pi/\lambda$, the eddy crossing time $\tau_{\rm eddy}
\sim \lambda/\Delta u$ (with $\Delta u$ being the turbulent velocity
fluctuation amplitude) becomes shorter than the cooling time
$\tau_{\rm cool} \sim 3 k_B T/(n \Lambda(T)) $, where $\Lambda(T)$ is
the interstellar cooling function. Although not really applicable
here, it is instructive to see that in case of incompressible
turbulence following a Kolmogoroff scaling law, where the energy
dissipation rate is given by $\varepsilon \sim
\rho {\Delta u}^3/\lambda$, we obtain a lower cut-off in wavelength,
where thermal instability becomes inhibited, if
\begin{eqnarray}
 \label{lambdacut}
  \lambda  &<&  \left(\frac{3 k_B \bar m}{\Lambda_0} \right)^{3/2} \, \varepsilon^{1/2} \, \frac{\temp^{3/4}}{\rho^2}\\
  &\approx & 1.42\times 10^{19} \, {\rm cm} \nonumber \,,
\end{eqnarray}
taking $\varepsilon \sim 10^{-26}\, {\rm erg} \, {\rm cm}^{-3}\,
{\rm s}^{-1}$ for SN energy injection; a simple cooling law for the
warm neutral medium of $\Lambda(T) = \Lambda_0 T^{1/2}$ has been
adopted, with $\Lambda_0 \approx 1.9 \times 10^{-27}\, {\rm erg} \,
{\rm cm}^3 \, {\rm s}^{-1} \, {\rm K}^{-1/2}$ (taken from the
cooling curve of Dalgarno \& McCray 1972) for a WNM of a density of
$n = 0.3 \, {\rm cm}^{-3}$, a temperature of $T=1000$ K, and a low
degree of ionization $x \approx 0.01$. Therefore rough numerical
estimates are typically of the order of parsecs, consistent with our
numerical resolution.  In fact, the critical wavelength $\lambda$
varies with temperature, degree of ionization and hence cooling; for
the WNM we find quite a large range of values from $10^{17} -
10^{20}$ cm, according to Eq.~(\ref{lambdacut}). We will investigate
the case of compressible, superalfvenic and supersonic turbulence in
more detail in a forthcoming paper.

A concern with the present simulations is the numerical resolution,
which is always a limitation, to which HD/MHD simulations are
subject. The crucial questions are (i) how do the results change as
the resolution is increased and (ii) how is the amount of cold gas
affected by a resolution larger than the Field length, $\lambda_{F}=
0.1$ pc in the WNM and 0.001 pc in the CNM (see e.g., Audit \&
Hennebelle 2005), far below the limit one can handle in global
simulations. Note that both Koyama \& Initsuka (2002) and Audit \&
Hennebelle (without magnetic fields) used 2D boxes of 0.3 pc and 20 pc
in length, three and two orders of magnitude smaller than our box size
and cell resolutions of the order of 0.01 parsec. The first of these
questions has been discussed at length for the HD run in Paper I,
where it has been shown that the large scale structure and global
properties of the ISM of the different thermal regimes remain
essentially unchanged, once a minimum resolution of 1.25 pc is
achieved. With regard to (ii) the calculation of the cooling length
depends also on heat conduction, which may be suppressed in stochastic
magnetic fields as the ones seen in the present simulations (for a
detailed discussion see Paper I; see also Malyshkin \& Kulsrud 2001). However,
as shown above in a dynamical ISM the Field criterion can be
superseded, as the cooling fluid element may be overturned by
turbulent fluctuations in the velocity and density field and
therefore, some of the gas that is heated up or cooling down to an
unstable regime may not participate in a transition to a stable phase
at lower temperature.

The present simulations still neglect an important component of the
ISM, i.e. high energy particles, which are known to be in rough
energy equipartition \textit{locally} with the magnetic field, the
thermal and the turbulent gas in the ISM. The presence of CRs and
magnetic fields in galactic halos is well known and documented by
many observations of synchrotron radiation generated by the relativistic electron
component. There is increasing evidence that the fraction of these
cosmic rays that dominates their total energy is of Galactic origin
and can be generated in supernova remnants via the so-called
diffusive shock acceleration mechanism to energies up to $10^{15}$
eV (for original papers see Krymski et al. 1977, Axford et al. 1977,
Bell 1978, Blandford \& Ostriker 1978, for a review Drury 1983, for
more recent calculations see Berezhko 1996). It is also common
wisdom that the propagation of these particles generates MHD waves
due to the so-called streaming instability (e.g.\ Kulsrud \& Pearce
1969) and thereby enhances the turbulence in the ISM. In addition,
self-excited MHD waves will lead to a dynamical coupling between the
cosmic rays and the outflowing fountain gas, which will enable part
of it to leave the galaxy as a galactic wind (Breitschwerdt et al.
1991, 1993, Dorfi \& Breitschwerdt 2005). Furthermore, as the cosmic
rays act as a weightless fluid, not subject to radiative cooling,
they can bulge out magnetic field lines through buoyancy forces.
Such an inflation of the field will inevitably lead to a Parker type
instability, and once it becomes nonlinear, it will break up the
field into a substantial component parallel to the flow (Kamaya et
al.\ 1996), thus facilitating gas outflow into the halo. It has been
suggested that eventually reconnection with an undisturbed halo
field component (Tanuma et al. 2003) and/or due to twisted
approaching flux tubes driven by Coriolis forces (Hanasz et al.
2002), will dissipate some of the field energy and change the field
topology. This may even provide the necessary ingredients for a
Galactic $\alpha \omega$-dynamo to operate, generating a large scale
poloidal field as was originally proposed by Parker (1992). It is
still a matter of debate, if Galactic halo conditions are suitable
for anomalous resistivity to occur in current sheets with a minute
thickness of the ion Larmor radius in order to promote fast
reconnection at about 10\% of the Alfv\'en speed. In any case will
such a scenario alleviate the transport of hot plasma into the halo.
The effect of cosmic rays on large scale ISM evolution will also be
the subject of a forthcoming paper.

\section{Summary and final remarks}

In the present paper we investigate the r\^ole of magnetic field and
matter circulation between the disk and halo and its effect on the
dynamical evolution of the disk gas. These results are compared to
HD simulations of the same resolution that we have carried out as
well. The main results of this paper can be summarized as follows:
\begin{itemize}

\item The simulations show the presence of a wiggly thin disk of cold gas
(both in density as well in magnetic field) overlayed by a frothy
thick disk composed of neutral and ionized hydrogen having different
scale heights.

\item The thin magnetized disk is overlayed by Parker-like loops (produced
without cosmic rays), magnetic islands, clouds wrapped in field lines
moving downwards and cold gas descending along the Parker loops, which
have risen to several kpc above the plane within 130 Myrs.

\item The highest density gas tends to be confined into shock
compressed layers that form in regions where several large scale
streams of convergent flow (driven by SNe) occur.

\item The compressed regions, which have on average lifetimes
of a few free-fall times (about 10-15 Myr), are filamentary in
structure, tend to be aligned with the local field and are
associated with the highest field strengths (in the MHD run), while
in the HD run there is no preferable orientation of the filaments.
The formation time of these high density structures depends on how
much mass is carried by the convergent flows, how strong the
compression and what the rate of cooling of the regions under
pressure are (see also Ballesteros-Paredes 2004).

\item The magnetic field has a high variability (which decreases towards
higher gas densities) and it is \emph{largely uncorrelated with the
density}. The field is driven by inertial motions (which is consistent
with the dominance of the ram pressure), rather than it being the
agent determining the gas motions. In the latter case the field would
not be strongly distorted, and it would direct the motions
predominantly along the field lines.

\item T$\leq 200$ K gas has $\mbox{P}_{B}>\mbox{P}_{ram}\gg\mbox{P}_{th}$,
demonstrating that magnetically dominated regions do exist. For $200
<\temp \leq 10^{5.5}$ K ram pressure determines the dynamics of the
flow, and therefore, the magnetic pressure does not act as a
significant restoring force. Near supernovae thermal and ram
pressures determine the dynamics of the flow. The hot gas in
contrast is controlled by the thermal pressure, since magnetic field
lines are swept towards the dense compressed walls.

\item The existence of ram pressure dominated flows leads inevitably to
shear flows, thus driving strong superalfvenic and supersonic turbulence
in the ISM.

\item The volume filling factors of the different ISM temperature regimes
depend sensitively on the existence of a duty cycle between the disk
and halo acting as a pressure release valve for the hot ($\temp >
10^{5.5}$ K) gas in the disk. The mean occupation fraction of the hot
phase in \emph{both HD and MHD runs} is $\sim 17-21\%$. The $\temp\leq
200$ K and $10^{3.9} <\temp\leq 10^{4.2}$ K regimes occupy $5-6\%$ and
$10-11\%$ of the disk volume, respectively, while the thermally
unstable regimes ($200 <\temp\leq 10^{3.9}$ and $10^{4.2} <\temp\leq
10^{5.5}$ K) fill in total up to 70\% of the disk volume in both runs.

\item Most of the disk mass is found in the $\temp\leq 10^{3.9}$ K gas,
with the cold ($\temp \leq 200$ K) and thermally unstable gases
($200<\temp\leq 10^{3.9}$ K) harbouring on average 83 and 93\% of the
disk mass in the MHD and HD runs. About 64-67\% of the WNM gas has
temperatures in the range $500-5000$ K in both runs.

\item With the magnetic field present and initially orientated
parallel to the disk varying as $\rho^{1/2}$, transport into the halo
is inhibited but not prevented. On larger scales magnetic tension
forces are weaker than on the smallest scales and therefore vertical
expansion still takes place efficiently and the occupation fraction of
the hot gas becomes comparable to the values observed in the
hydrodynamical simulations. Thus, hot gas is fed into the Galactic fountain at almost a similar rate than without the field.

\end{itemize}

In the dynamical picture of the ISM emerging from our 3D
high-resolution simulations, turbulence generated by shear flows of
expanding bubbles seems to be a key element of structure formation in
both HD and MHD runs. Starting from here and following a bottom-up
scheme we feel encouraged to study in the future other physical processes and
ingredients in more detail, such as non-equilibrium cooling,
self-gravity, heat conduction and cosmic rays, respectively, to name
the most important ones.

\begin{acknowledgements}
M.A. and D.B. are partially supported by the ESO/FCT (Portuguese
Science foundation) grant PESO/P/PRO/40149/2000. D.B. thanks G.
Hensler for useful discussions on the subject. M.A. benifited from
discussions with E. Vazquez-Semadeni and M.-M. Mac Low.
\end{acknowledgements}

\end{document}